\documentclass{pasj00}

\SetRunningHead{Oda et al.}{Cosmic Supernova Rate Density Evolution}

\begin{document}
\title{Implications for Galaxy Evolution from the Cosmic Evolution of
Supernova Rate Density} 

\author{Takeshi \textsc{Oda} \altaffilmark{1}, 
Tomonori \textsc{Totani} \altaffilmark{1}, 
Naoki \textsc{Yasuda} \altaffilmark{2}, 
Takahiro \textsc{Sumi} \altaffilmark{3}, \\
Tomoki \textsc{Morokuma} \altaffilmark{5},
Mamoru \textsc{Doi} \altaffilmark{4},
George \textsc{Kosugi} \altaffilmark{5}
}

\altaffiltext{1}{Department of Astronomy, School of Science, 
Kyoto University, Sakyo-ku, Kyoto 606-8502, Japan} 
\email{takeshi@kusastro.kyoto-u.ac.jp}

\altaffiltext{2}{Institute for Cosmic Ray Research, University
of Tokyo, Kashiwa, Chiba, 227-8582, Japan}

\altaffiltext{3}{Solar-Terrestrial Environment Laboratory, Nagoya
University, Furo-cho, Chikusa-ku, Nagoya, 464-8601, Japan}

\altaffiltext{4}{Institute of Astronomy, School of Science, 
University of Tokyo, 2-21-1 Osawa, Mitaka, Tokyo, 181-0015, Japan}

\altaffiltext{5}{National Astronomical Observatory of Japan, 2-21-1
Osawa, Mitaka, Tokyo, 181-8588, Japan}

\KeyWords
{supernovae:general --- galaxies:evolution --- cosmology:observations}
\maketitle

\begin{abstract}
 We report a comprehensive statistical analysis of the observational
 data of the cosmic evolution of supernova (SN) rate density, to derive
 constraints on cosmic star formation history and the nature of type Ia
 supernova (SN Ia) progenitor. We use all available information of
 magnitude, SN type, and redshift information of both type Ia and
 core-collapse (CC) SNe in GOODS and SDF, as well as SN Ia rate
 densities reported in the literature. Furthermore, we also add 157 SN
 candidates in the past Subaru/Suprime-Cam data that are newly reported
 here, to increase the statistics. We find that the current data set of
 SN rate density evolution already gives a meaningful constraint on the
 evolution of the cosmic star formation rate (SFR) at $z \lesssim 1$,
 though strong constraints cannot be derived for the delay time
 distribution (DTD) of SNe Ia. We derive a constraint of $\alpha \sim $
 3--4 [the evolutionary index of SFR density $\propto (1+z)^\alpha$ at
 $z \lesssim 1$] with an evidence for a significant evolution of mean
 extinction of CC SNe [$E(B-V) \sim 0.5$ at $z \sim 0.5$ compared with
 $\sim 0.2$ at $z=0$], which does not change significantly within a
 reasonable range of various DTD models. This result is nicely
 consistent with the systematic trend of $\alpha$ estimates based on
 galactic SFR indicators in different wavelengths (ultraviolet,
 H$\alpha$, and infrared), indicating that there is a strong evolution
 in mean extinction of star forming regions in galaxies at relatively
 low redshift range of $z \lesssim 0.5$.  These results are obtained by
 a method that is completely independent of galaxy surveys, and
 especially, there is no detection limit about the host galaxy
 luminosity in our analysis, giving a strong constraint on the star
 formation activity in high-$z$ dwarf galaxies or intergalactic space.
\end{abstract}

\section{INTRODUCTION}

In recent years a number of searches for high redshift supernovae (SNe)
have been conducted. Although the primary purpose of most of these
surveys is measurement of the cosmic expansion, these surveys also
allowed measurements of the cosmic supernova rate density and its
evolution \citep{Pain02, Tonry03, Madgwick03, Gal-Yam04, Blanc04,
Maoz04, Dahlen04, Cappellaro05, Barris06, Neill06, Poznanski07,
Sharon07, Mannucci07, Kuznetsova07}. Studying these data should provide
us with important information not only for the cosmic star formation
history (CSFH) but also the still unknown progenitor of type Ia
supernovae (SNe Ia).
The progenitor of SNe Ia is believed to be a binary system including a
white dwarf, and the SN Ia rate density evolution is a convolution of
the cosmic star formation history and the delay time distribution (DTD)
from star formation to SN Ia events.  DTD depends on the progenitor
models, and hence to constrain DTD observationally is a useful approach
to reveal the SN Ia progenitor (Madau et al. 1998; Yungelson \& Livio
1998, 2000;  Dahlen \& Fransson 1999; Gilliland et al. 1999; Gal-Yam \&
Maoz 2004; Strolger et al. 2004, 2005;  Barris et al. 2004; Oda \&
Totani 2005, hereafter OT05; Strigari et al. 2005).

However, it is not an easy task to actually extract useful constraints
from the SN rate density evolution data. Previous studies \citep{Maoz04,
Strolger04, Forster06} mainly concentrated on the determination of DTD,
using the rate density evolution of SNe Ia.  In such an analysis,
sometimes CSFH models are assumed based on the observational estimates
from high-$z$ galaxy surveys. However, as argued by \citet{Forster06},
the constraint on DTD models sensitively depends on the assumed CSFH,
and hence it is difficult to derive a robust constraint on DTD.

The primary purpose of this paper is to perform a comprehensive
likelihood analysis using all available SN rate density evolution data
in the literature, to derive constraints on DTD and/or CSFH.  After the
GOODS high-$z$ supernova survey \citep{Dahlen04, Strolger04}, whose data
was used in \citet{Strolger04} and \citet{Forster06}, a number of
observational estimates of SN rate density evolution have been published
(mostly for SNe Ia, but some data also for CC SNe). Our approach is to
derive constraints only by using SN rate data, without using information
of CSFH from galaxy surveys. We will perform a simultaneous fit to both
the SN Ia and CC SN rate density evolution data, surveying parameters of
the CSFH model with a variety of DTD models. We will find that, though a
strong constraint on DTD models cannot be derived even from all the
available data so far, we can set interesting constraints on CSFH and
evolution of mean dust extinction of CC SNe, which can be compared with
those inferred from galaxy surveys.

Although there are a number of observational estimates on CSFH at a
variety of redshifts from galaxy surveys, there is still a large
uncertainty in the star formation rate (SFR) density estimated from
galaxy observations, because of extinction, initial mass function, or
extrapolation of luminosity functions to fainter magnitudes below the
detection limits (see Hopkins 2004 and Hopkins \& Beacom 2006, and
reference therein). Therefore it is useful and important to derive
constraints on CSFH from SNe independently of galaxy surveys.  In
contrast to SFR density estimates by galaxies, detectability of SNe does
not depend on the host galaxy brightness, and even intergalactic star
formation activity can be probed by hostless SNe. Searches for $z \sim
1$ SNe are typically performed at wavelength around the $i'$ and $z'$
band roughly corresponding to the rest-frame visual bands, and hence the
effect of extinction by dust is expected to be smaller than the CSFH
estimates based on the rest-frame UV emission of galaxies. It is not
trivial that a unit mass of star formation always produces the same
number of SNe, but it could evolve with redshift or physical properties
of galaxies.  If a significant difference between CSFH inferred from
galaxy surveys and that from SN surveys is found, it might indicate that the
relation between star formation and supernova production is not as
simple as normally assumed.

In addition to the available SN rate density data in the literature, we
also utilize the photometric sample of SN candidates found in the past
observations using Subaru/Sprime-Cam. This Subaru Supernova Survey (SSS)
sample includes 157 supernova candidates, 61 out of which have clear
offsets from the centers of host galaxies and hence they are most likely
SNe. This data set is complementary to GOODS, SNLS and the IfA deep
survey \citep{Strolger04,Neill06,Barris06} in terms of the combination
of the survey area and depth; the covered area of SSS, $1.4$ deg$^2$, is
wider than the GOODS, and the SSS depth, $i' \sim 26.0$, is deeper than
the SNLS and the IfA deep survey. Though no SN type or redshift
information is available for the SSS sample, we add this data set to our
likelihood analysis to increase the statistics especially for CC
SNe. Compared with SNe Ia, there are not many data of the rate density
for CC SNe. Combined analysis of the SSS counts including all SNe and
other data for SN Ia rate density evolution should give some constraints
on the CC SN rate density evolution and hence CSFH.

The following are the plan of this paper. In \S \ref{sec:data}
and \S \ref{sec:obs_res} we describe the SSS data set and 
analysis procedure of selecting SN candidates. Formulations
of the comparison of the theoretical model and the observational
data are given in \S \ref{sec:comp_method}. 
Constraints on the CSFH from our comprehensive 
parameter survey are derived in \S \ref{sec:results}. 
Conclusions are given in \S \ref{sec:summary}. Throughout
this paper, the standard $\Lambda$CDM universe is assumed with the
following values of the cosmological parameters: $\Omega_{\rm M}$ = 0.3, 
$\Omega_{\rm \Lambda}$ = 0.7, $h_{70} \equiv H_{0}$/ (70 km s$^{-1}$
Mpc$^{-1}$) = 1. All magnitudes are given in the AB magnitude system.

\section{The SSS Data}\label{sec:data}
The SSS data set consists of 
the following three fields, named A2152, MS1520.1, and
the spring field (SF), whose positions on the sky are given in Table 
\ref{table:fields}. All images are taken with the Subaru/Suprime-Cam
\citep{Miyazaki02} having an effective field-of-view (FOV) of 
$30' \times 24'$, with a time interval of about one month that is
suitable for a high-z supernova search. 
We describe details of the observations at each field below.

\textit{A2152 field} -- A single FOV of the Suprime-Cam centered on 
the galaxy cluster Abell 2152 was observed, where two galaxy clusters
(A2152 at $z = 0.04$ and A2152-B at $z = 0.13$) closely overlap on the 
line of sight \citep{Blakeslee01}. 
About one month after the first imaging of this field
(2003 May 5), the field was imaged again during four consecutive nights
(June 1--4). Images were taken with $V_{c}$ and $I_{c}$ band filters and
typical exposure time is a few hours for each filter per day, but we use
only the $I_{c}$ band images for our supernova search. About 40 \% of
supernova candidates found in $I_{c}$ band data of this field were not
detected on $V_{c}$ band images, while there is no SN candidate that was
detected only in $V_{c}$ band. 

\textit{MS1520.1 field} -- A single FOV of the Suprime-Cam was
observed around the galaxy cluster MS1520.1+3002 at $z = 0.117$
\citep{Stocke91}. Observations were performed on April 25 and May 20 in
2001 with the $i'$ band filter. The exposure time is about one hour. 

It should be noted that the expected number of supernovae in the galaxy
clusters in the MS 1520.1 and A2152 fields is too small to affect the
conclusions of this paper (Gal-Yam et al. 2002; Sharon et al. 2007), and
hence the existence of these clusters is not taken into account in our
theoretical modeling. 

\textit{The spring field} -- There are four adjacent Suprime-Cam images
of this field in $i'$ band, which we call SF 1-4.  The first images were
taken on March 19, and the second and third ones were on April 9 and 11,
in 2002. Typical exposure time of each field is about one hour, but the
observational conditions of SF2 and SF3 are better than those of SF1 and
SF4.  This field and the MS1520.1 field were observed as a part of the
Supernova Cosmology Project. Thus, the observation was designed to find
high redshift SNe Ia for the cosmological purpose, and follow-up
spectroscopic observations of some supernova candidates were
performed. As a result, three SNe (2002fc, 2002fd and 2002fe) are
clearly identified to be SNe Ia, and another SN (2002ff) is a possible
candidate of SN Ia. Their redshifts are 0.88, 0.278, 1.086, and 1.1,
respectively.  (See IAU circ. 7971 for more details.) All of these are
included in our SN candidates, but the information obtained with
spectroscopic follow-up is not used, because the majority of the SNe in
the SSS sample do not have spectroscopic information, and adding these
spectroscopic information hardly affects the conclusions of this paper.

\begin{table*}
\begin{center}
\caption{ Basic Information of the SSS Observations
\label{table:fields}}
\begin{tabular}{ccccccc}
\hline
Field name & R.A. & Decl. & Area &  Observing dates & 
Typical exposure time  & Band filter \\
 & & &  [deg$^2$] & & (hour) &\\
\hline
A2152 & \timeform{16h05m22s} & \timeform{+16D26'55''} & 0.21 & 2003
 May 5, June 1-4 & 3 & $I_{C}$, $V_{C}$ \\ 
MS1520.1 & \timeform{15h22m13s} & \timeform{+29D51'59''} & 0.23 &
 2001 April 25, May 20 & 1 & $i'$ \\
SF 1 & \timeform{14h00m56s} & \timeform{+05D40'48''} & 0.24 & 2002
 March 19, April 9, 11 & 1 & $i'$ \\
SF 2 & \timeform{13h58m36s} & \timeform{+05D22'30''} & 0.23 & 2002
 March 19, April 9, 11 & 1 & $i'$ \\
SF 3 & \timeform{14h03m46s} & \timeform{+05D11'04''} & 0.23 & 2002
 March 19, April 9, 11 & 1 & $i'$ \\ 
SF 4 & \timeform{14h13m18s} & \timeform{+05D40'43''} & 0.24 & 2002
 March 19, April 9, 11 & 1 & $i'$ \\
\hline
\end{tabular}
\end{center}
\end{table*}

\section{Selection of the SSS Supernova Candidates}\label{sec:obs_res}

\begin{figure*}
 \begin{center}
  \FigureFile(160mm, 80mm){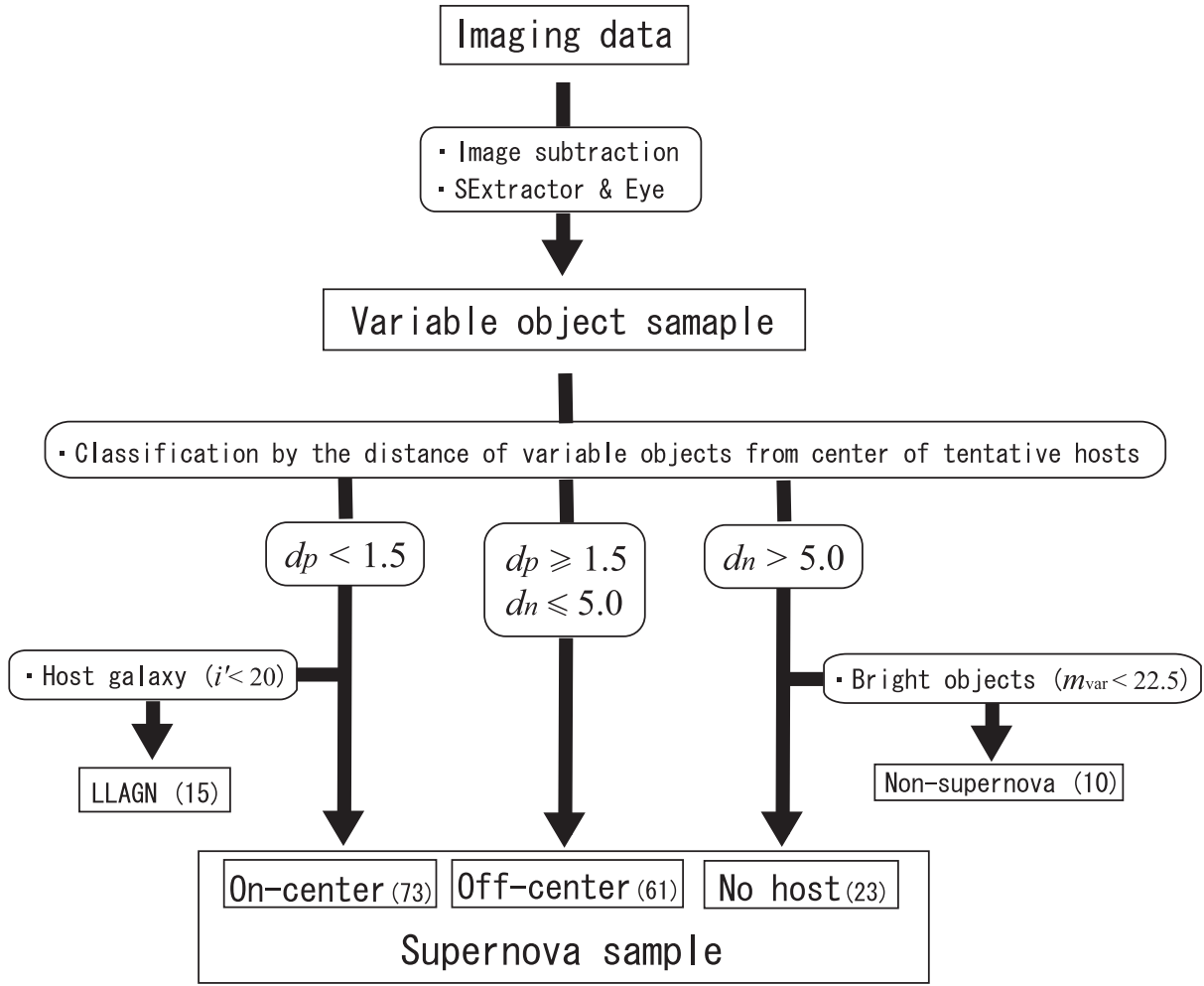}
 \end{center}
 \caption{ A flow-chart showing the detection and selection
 procedures of the supernova candidates in SSS.
\label{fig:flowchart}}
\end{figure*}

Here we describe how we selected supernova candidates from
the SSS data in detail, and a schematic flow-chart of
these processes is presented in Fig. \ref{fig:flowchart}.

\subsection{Variable Object Detection, Detection Efficiency and 
Position Accuracy}
\label{sec:procedures}

First, we made differential images from image pairs separated by about
one month, using the image subtraction method ISIS \citep{Alard98,
Alard00}.  Source detection was carried out using the SExtractor
software \citep{Bertin96}. We detected candidate variable objects
requiring that they have five or more connected pixels whose counts are
more than 1$\sigma$ level of the surface brightness fluctuation, after
\timeform{0''.7} FWHM Gaussian smoothing on the subtracted images.  The
variability magnitude ($m_{\rm var}$) corresponding to the flux on the
subtracted image was measured by the SExtractor's automatic aperture
magnitude. From these candidates, objects having high signal-to-noise
(S/N) were selected. The criteria are S/N = 7 - 10, which depend on
observational conditions of the fields. Finally we checked the images of
these objects on the subtracted and original frames by eye in order to
eliminate spurious objects.

The detection efficiency $\varepsilon (m_{\rm var})$ and the position
accuracy of variable objects on the subtracted images are estimated by
simulations using artificial point sources placed randomly on one of the
pre-subtraction images, under the exactly same object selection criteria
as described above.  Ideally, this test should be performed on various
background conditions (e.g., in the blank field or on a host galaxy),
since the detection efficiency could be changed by the location of
variable sources. However, we ignore this effect in this paper, because
the typical surface brightness of supernova host galaxies at $z \gtrsim
0.5$ is fainter than the sky level, and hence the noise of image is
dominated by the sky background.  In fact, we confirmed that there is no
marked difference in the fluctuation of photon counts of the subtracted
image between the blank field and locations of galaxies having typical
magnitudes of supernova hosts. This result is in agreement with previous
studies (e.g., Strolger et al. 2004; Poznanski et al. 2007).  Following
\citet{Strolger04}, detection efficiency estimated for various
magnitudes of flux variability is fitted by the following function:
\begin{equation}
\label{eq:fitting}
\varepsilon (m_{\rm var}) = \frac{1}{1 + \exp[ (m_{\rm var} -
m_{0})/S_{\rm fit} ]}.
\end{equation}
We use a single value of $S_{\rm fit}=0.43$ for all fields, but
different values of $m_0$ are fitted to the simulations in different
fields. The fitting results of $m_0$ are given in Table
\ref{table:num_counts}.

\begin{figure*}
 \begin{center}
  \FigureFile(160mm, 45mm){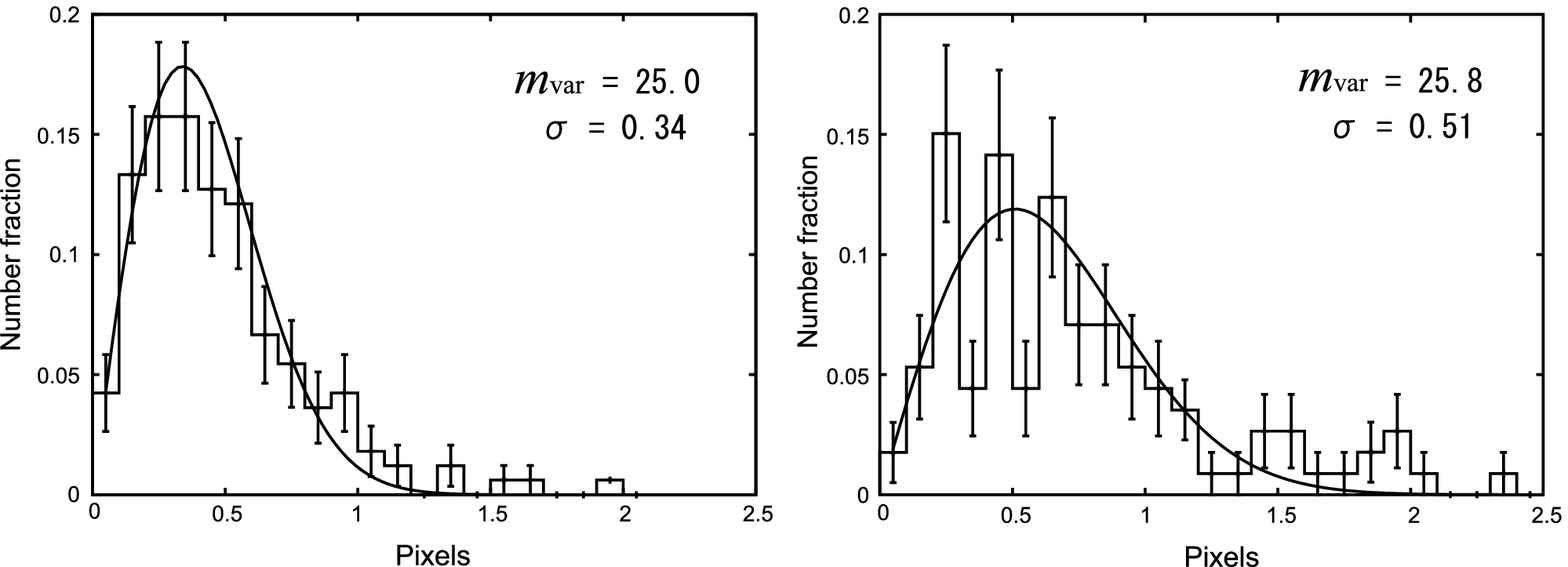}
 \end{center}
 \caption{ Histograms of the
 distance between original and recovered positions of artificial point
 sources in the A2152 field. 
 Left and right panels show the cases of two different magnitudes
 of the artificial sources ($m_{\rm var} = 25.0$ and 25.8),
 respectively. Error bars show a 1$\sigma$ statistical error.  Solid
 lines are fits by two-dimensional Gaussian distribution 
 [$\propto r \exp (-r^2/2 \sigma^2)$] with $\sigma = 0.34$ and 0.51
 pixel, respectively. 
 \label{fig:posi_art}}
\end{figure*}

We estimate the accuracy of position recovery by measuring the
positional offsets of detected objects from the original positions.  As
shown in Fig. \ref{fig:posi_art}, the offsets well obeys the
two-dimensional Gaussian distribution.  Their standard deviations in one
dimension derived by least square fits are $\sigma = 0.34$ and 0.51
pixel for $m_{\rm var} = 25.0$ and 25.8, respectively, in the A2152
field (1 pixel = \timeform{0''.2}).

\subsection{Host Galaxies}
\label{sec:hostGal}
Many extragalactic variable objects should be associated with host
galaxies, and their nature and positional relation to variable objects
are important information to select supernovae. First we simply define a
tentative host galaxy as the object in the reference image whose surface
brightness peak is the closest to a detected variable object. Sometimes
the host galaxies are classified as point sources (the SExtractor
stellarity parameter greater than 0.8), and their positions are the same
as those of variable point sources.  In such cases they could be simply
the variable sources such as QSOs or variable
stars. Furthermore, we cannot exclude a possible contribution from
unresolved host galaxies. In these cases, we assign the next closest
object in the reference frame as the tentative host galaxies. Therefore,
positions of the variable objects are always different from the
centers of their tentative host galaxies by definition.
 
Based on the estimates of the position accuracy for variable objects, we
call an object ``on-center'' when $d_p <$ 1.5, where $d_p$ is the
distance from the center of host galaxies to variable objects measured
in units of pixel.  The center of host galaxies is
simply defined by the surface brightness peak.  About 97 \% and 85 \% of
the bright ($m_{\rm var} = 25.0$) and faint ($m_{\rm var} =$ 25.8) point
sources are found within 1.5 pixels (= 0''.3) from the original
position, respectively, according to the simulation using artificial
sources in the A2152 field. 

We define the distance between a variable object and its host galaxy
center that is normalized by the size of the host galaxy, as $d_n \equiv
d_p / r_{p, \rm gal}$, where $r_{p, \rm gal}$ is the effective size of
the host galaxy, defined as the radius of the ellipse from the host
center to the direction of the variable object.  The ellipse is obtained
by the fitting to the host galaxy, as calculated in the SExtractor, and
its size is determined so that the squares of the major and minor axes
are the same as the second order moments along the axes, i.e.,
\begin{eqnarray}
a_j^2 \equiv \frac{\sum_{i \in S} F_i x_{i, j}^2}{\sum_{i \in S} F_i} \ ,
\end{eqnarray}
where the subscript $i$ denotes for each pixel, $F_i$ the flux counts in
a pixel in $I_c$- (A2152 field) or $i'$-band (other fields), $x_{i, j}$
is the distance from the host center to the $i$-th pixel projected onto
the major or minor axis (denoted by the subscript $j$), and the
summation is over the whole region of the host galaxy.

\begin{figure}
 \begin{center}
  \FigureFile(80mm, 45mm){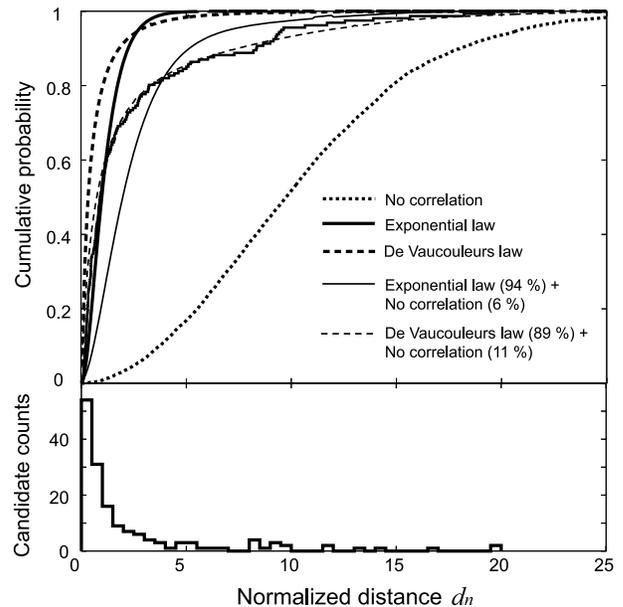}
 \end{center}
 \caption{
 Cumulative (top) and differential (bottom)
 distributions of the normalized distance $d_n$
 between the positions of variable objects and the centers of host
 galaxies. Model distributions derived from the exponential and de
 Vaucouleurs laws are shown with thick solid and dashed lines,
 respectively. The distribution expected for objects located randomly on
 the reference images is shown with the dotted line. The thin
 solid/dashed lines are the combined distribution of the exponential/de
 Vaucouleurs law and the random distribution. The relative proportions
 between the different components are indicated in the figure. 
 \label{fig:distance}}
\end{figure}

Now we examine the $d_n$ distribution of the tentative host galaxies,
which is shown in Fig. \ref{fig:distance}. The distribution clearly
shows a stronger correlation between variable objects and host galaxies
than that expected for objects that are randomly distributed on the sky,
indicating that the majority of variable objects are physically
associated with galaxies. The radial distribution of supernovae in their
host galaxies is still uncertain (e.g., Bartunov et al. 1992; Howell et
al. 2000), and here we test the galaxy surface brightness profiles often
used in the literature, i.e., the exponential and the de Vaucouleurs
profile. Here, we have taken into account the effect of seeing for the
surface brightness profile, by relating the effective radius of the
original profiles to the seeing-convolved second order moments. 
It should be noted that the simple exponential or de Vaucouleurs
law may not be sufficient to describe all galaxies; there may be
contribution from irregular galaxies, and cosmological surface
brightness dimming effect may alter significantly the apparent
profile (e.g., Totani \& Yoshii 2000).
However these effects are difficult to model quantitatively, and
they are ignored here for the simplicity.

We find that the observed distribution is different from what expected
when all variable objects obey the exponential or the de Vaucouleurs
profile. However, the distribution is well described by the combination
of the two components: a galaxy profile (exponential or de
Vaucouleurs) and a random distribution.  We find the best-fit relative
proportion by the Kolmogorov-Smirnov test as 94 (88)\% for the
exponential (de Vaucouleurs) profile and the rest for a random
distribution. The de Vaucouleurs profile gives an especially good fit to
the observed distribution.  However, it does not necessarily mean that
the de Vaucouleurs profile is better than the exponential for SN
distribution, since there may be a significant contribution from AGNs.
From this figure, we find that almost all objects with
$d_n > 5$ are likely to be unrelated to the tentatively assigned host
objects, and hence we define objects with $d_n > 5$ as those without
detectable host galaxies. 

\begin{table*}
\begin{center}
\caption{ Depth of surveys and number of detected SN candidates.
\label{table:num_counts}
}
\begin{tabular}{ccccccc}
\hline
 &  &  & 
\multicolumn{4}{c}{Number of SN candidates} \\
\cline{4-7}
Field & $m_0$ [mag] & & All & On-center &
 Off-center & No host \\
\hline
A2152 & 25.8 & & 33 & 12 & 17 & 4 \\
MS1520.1 & 26.1 & & 13 & 5 & 7 & 1\\
SF 1 & 25.3 & & 20 & 11 & 5 & 4 \\
SF 2 & 26.0 & & 30 & 15 & 12 & 3 \\
SF 3 & 26.0 & & 41 & 20 & 15 & 6 \\
SF 4 & 25.6 & & 20 & 10 & 5 & 5 \\
Total & & & 157 & 73 & 61 & 23 \\
\hline
\end{tabular}
\end{center}
\end{table*}

\subsection{The Supernova Candidates}

\subsubsection{Off-center Supernova Candidates}

Now we have a robust sample of supernovae, i.e., the 61 variable objects
associated with host galaxies and their locations are off-center on the
host galaxies ($d_{\rm p} > 1.5$ pix and $d_n \leq 5$).  Because of
these properties, the majority of them should be supernovae rather than
AGNs. A possible contamination is chance superpositions of background
AGNs in front of unrelated foreground galaxies (Gal-Yam et al. 2007).
We can make a rough estimate of such events as follows.  From the
statistics of a similar variable object search by Morokuma et al. (2007)
using the Subaru XMM-Newton Deep Survey (SXDS) data set, about 40 AGNs
are expected in the SSS data. The surface area covered by galaxies with
a similar magnitude to that of host galaxies in the SSS is about 5\% of
the total survey area, and hence we expect a few random
superpositions. This number is much smaller than the off-center
supernova candidates, and hence this effect can be neglected.

\subsubsection{On-center Supernova Candidates}
\label{sec:on-center} When variable objects are on the center of their
host galaxies, we cannot discriminate between the two possibilities of
supernovae or AGNs. However, some of these objects show very faint
variability flux compared with the total magnitude of the host galaxies,
and these objects are most likely to be low-luminosity AGNs (LLAGNs)
with very low accretion rate as reported in \citet{Totani05}, because
supernovae are generally as bright as the brightest class of galaxies.
In fact, we found no off-center SN candidates associated with galaxies
brighter than $i' = 20$.  Therefore we removed 15 variable objects that
are located at the center of very bright galaxies ($i' < 20$).

The remaining 73 on-center variable objects are then called as
``on-center supernova candidates'', though we cannot exclude a
contamination of AGNs in this sample. However, if we assume that
supernovae trace the light of the host galaxies, we can estimate the
expected number of on-center supernovae from the number of off-center
supernovae, by extrapolation of a surface brightness profile.  We find
that 32 and 129 on-center supernovae are expected for the exponential
and de Vaucouleurs profiles, respectively. The reality is likely between
the two, and if we assume the S\'{e}rsic profile, we find that the
expected number becomes the same as the observed number with the
S\'{e}rsic's index of $n_{\rm ser} \sim 3$. These results indicate that
at least about half of the on-center objects are supernovae
(corresponding to the exponential profile).  In fact, as mentioned in
the previous subsection, we expect about 40 AGNs in the SSS from the
statistics of the SXDS variable object search (Morokuma et al. 2007),
i.e., about a half of the on-center candidates.

The effect of AGN contamination in the on-center sample will be examined
when we will compare the theoretical model of supernova rate evolution
to the observed data.

\subsubsection{No-host Supernova Candidates}
\label{sec:no-host-SNC} The variable objects without host galaxies
should also be examined since supernovae may be included in them. First
we notice that there are objects that are clearly much brighter than
expected for supernovae.  In the off-center supernova sample, there is
no object brighter than $m_{\rm var} = 22.5$ in the variability
magnitude. However, 10 objects in the no-host sample are brighter than
this magnitude in spite of the no-detection of a host. These objects are
most likely variable quasars or variable stars in our Galaxy, and hence
they are rejected from the supernova candidates.

Then, the remaining 23 objects are defined as the no-host supernova
candidates without detectable host galaxies, and there is no marked
difference between the variability magnitude distributions of this
sample and the off-center supernovae.  Thus, although we cannot exclude
significant contamination from quasars and Galactic variable stars, most
of these are possibly supernovae with host galaxies that are fainter
than the detection threshold ($i' = 25.0$) or truly intergalactic
supernovae. To examine the former possibility, we estimate $\eta(z)$,
which is the fraction of supernovae in host galaxies that are detectable
by SSS. We assume that the supernova rate in a galaxy is proportional to
the rest-frame $V$-band luminosity of a host galaxy.  This is an
assumption that should not be accurately correct;
CC SNe are expected to trace galactic light in shorter
wavelength such as rest-frame UV, and SNe Ia with a long delay time
would trace longer wavelength light that is related to the stellar mass.
However, our data set is limited about available band filters,
and we make this assumption for the present data set.

Then we can estimate $\eta(z)$ by the $V$-band luminosity function of
galaxies at a given $z$ and the SSS detection limit for galaxies.  We
assume the following values and redshift evolution of the Schechter
parameters of the luminosity function: $\alpha = -1.15 - \log(1+z)$ and
$M_* = -20.5 - 5.0 \ \log (1+z)$, from the observations by
\citet{Ilbert05}. The K-correction between the observed band ($i'$ or
$I$) and the rest-frame $V$ are calculated assuming the Sbc galaxy
template.  The calculated correction factors in this way is $\eta = $
0.90, 0.79 and 0.72 for $z = $ 0.5, 0.8, and 1.0, respectively.  The
typical redshifts of SNe that should be detectable in SSS is 
$\sim$0.5 and 1.0
for CC SNe and SNe Ia, respectively, and we detected 134 supernova
candidates with detectable host galaxies.  Therefore, most or perhaps
all of the 23 no-host candidates can be explained by those associated
with galaxies under the detection limit. In other words, there is no
evidence for a significant population of intergalactic supernova
population.

\subsection{Summary of SN Candidate Selections}

\begin{figure*}
 \begin{center}
  \FigureFile(160mm,70mm){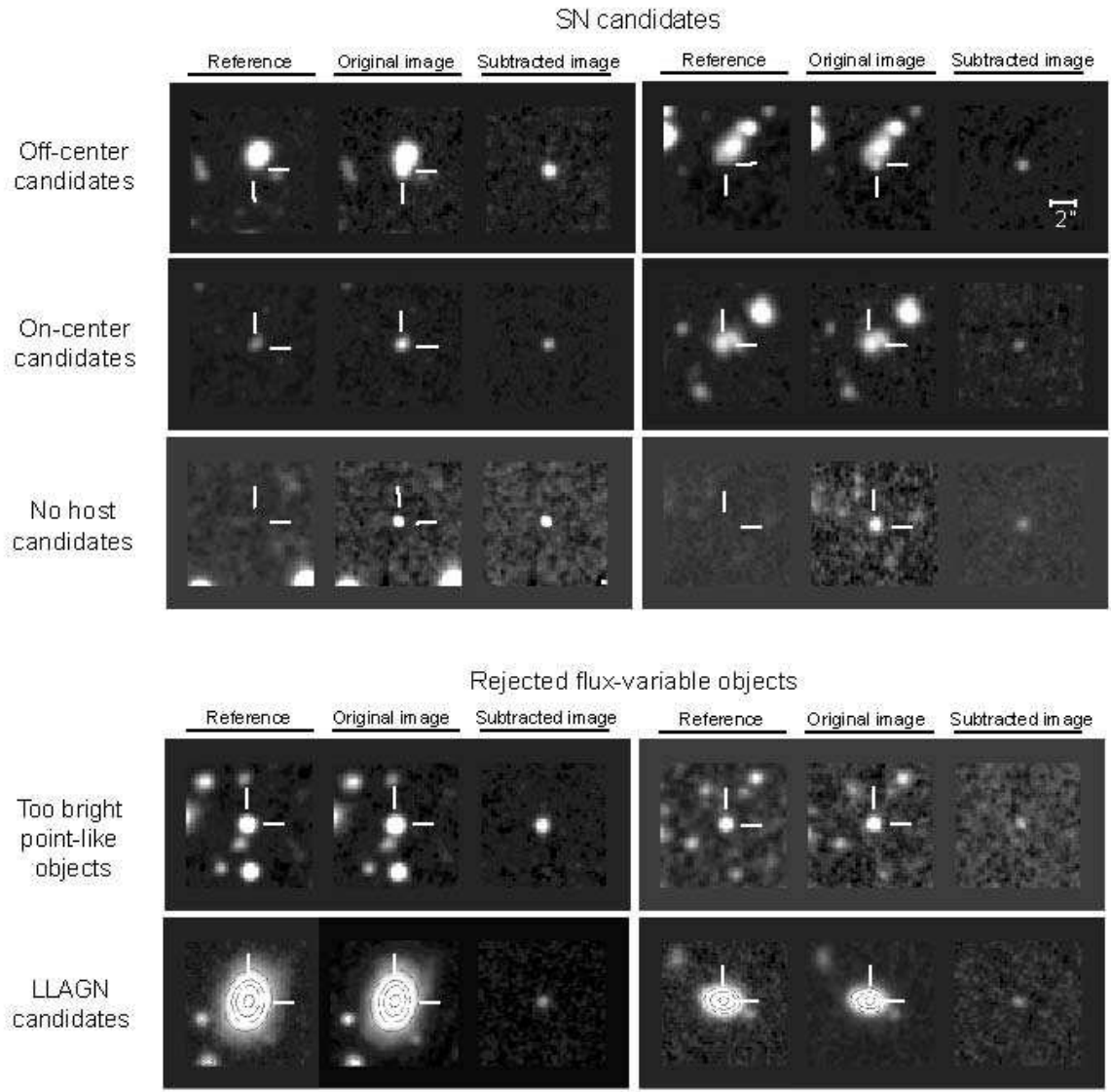}
 \end{center}
 \caption{ Images of the supernova candidates as well as rejected
 objects (see text).  In each panel, the left image is the reference
 (first epoch), the central image is the second epoch image, and the
 right image is the subtracted image. The positions of variabilities are
 indicated as the crossing points of two white bars. The scale of this image
 is shown in the upper right panel. 
\label{fig:images}}
\end{figure*}

\begin{figure*}
 \begin{center}
  \FigureFile(160mm,70mm){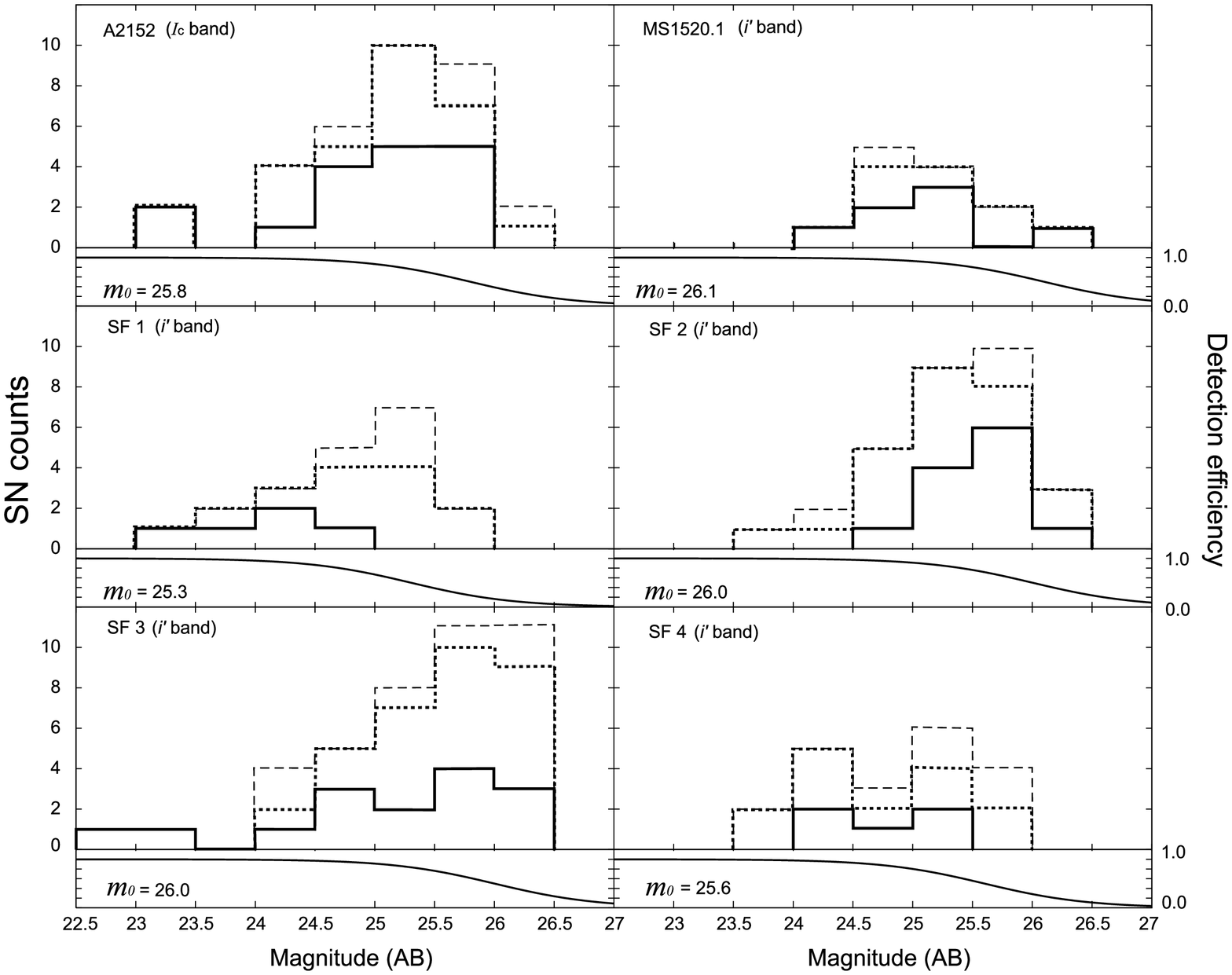} 
 \end{center}
 \caption{Variability magnitude ($i'$ or $I_c$)
  distributions of the supernova
 candidates detected in the six different SSS fields.  
  The distributions of the off-center SN sample (solid
 line), the off-center plus on-center samples (dotted line), and all the
 samples including the no-host sample are shown.  Solid curves in the
 small bottom panels show the detection efficiency in each field.
 \label{fig:SNhist}}
\end{figure*}

As a result of the above selection procedures, we find 157 supernova
candidates in total, including 73 on-center, 61 off-center, and 23
no-host candidates. Images of representative objects of these
classification are given in Fig. \ref{fig:images}, as well as the images
of those classified as non-SNe objects.  A summary of the results for each
field is presented in Table \ref{table:num_counts}.
The distribution of the variability magnitude of these supernova
candidates is shown in Fig. \ref{fig:SNhist}, as well as the estimated
detection efficiency.  The behavior of the faint end of the distribution
is in reasonable agreement with the detection efficiency estimate.  No
considerable field-to-field variation is found.

\subsection{The Sample Used for the Statistical Analysis}
\label{sec:SNsamples}

In the following likelihood analysis including the SSS data, we will
present two cases of (i) using all of the on- and off-center SSS samples
and (ii) using only the off-center SSS sample.  In the former case, all
on-center and off-center SN candidates are assumed to be the real
supernovae. In the latter case, the effect of the removal of central
regions of host galaxies is corrected assuming that the supernova
distribution obeys the exponential profile as calculated in \S
\ref{sec:on-center}. For this correction we simply multiplied a factor
of 0.65 to the theoretical prediction of the cosmic SN rate density,
since the mean fraction of light in central regions of all host galaxies
in the SSS is 0.35.  As described in \S \ref{sec:on-center}, the
exponential profile corresponds to assuming that about half of on-center
candidates are the real supernovae. Since the exponential profile is the
least concentrated one among the various profiles assumed in the
literature, the reality should be between the above two cases and hence
we can check the systematic uncertainty about the AGN contamination by
this treatment.

We do not include the no-host supernova candidates in the following
analysis, since we cannot exclude the contamination by variable quasars
or Galactic variable stars in this sample.  We have already shown in \S
\ref{sec:no-host-SNC} that the number of no-host candidates is similar
to that expected by SNe with host galaxies under the detection limit,
and hence there is no evidence for a significant population of truly
intergalactic supernovae. Therefore, we only make a
correction for SNe that are classified as no-host because their host
galaxies are fainter than the detection limit, by using the quantity
$\eta(z)$ calculated in \S \ref{sec:no-host-SNC}. This correction
factor is not far from unity, and it does not significantly affect our
conclusions even if this effect is completely ignored.

\section{Theoretical Model of SN Rate Evolution}
\label{sec:comp_method}
\subsection{ Cosmic Star Formation History }

For the parametrization of CSFH, we use the following
functional form (Gal-Yam \& Maoz 2004):
\begin{equation}\label{eq:SFH}
\Phi (z) \propto \left[ \left( \frac{1+z_{\rm break}}{1+z} \right)^{5\alpha} +
 \left( \frac{1+z_{\rm break}}{1+z} \right)^{5\beta} 
\right] ^{-0.2}.
\end{equation}
Here, $\alpha$ and $\beta$ are indices of the CSFH at low and high
redshift, respectively.  The CC SN rate evolution is simply assumed to
be proportional to the CSFH, because of their short life. The SN Ia rate
density $r_{\rm Ia}(z)$ is calculated from the CSFH convolved with the
DTD, $f_{D}(t_{\rm Ia})$, where the delay time $t_{\rm Ia}$ is elapsed
from star formation to the SN Ia events. In this paper, the parameter
$\alpha$ is treated as a free parameter that we constrain, while other
two parameters are fixed at $\beta = 0$ and $z_{\rm break} = 1.5$ in
our baseline model, in order to reduce the number of free
parameters. Although SNe Ia have a time delay from star formation, most
DTD models have the peak of the distribution at relatively small $t_{\rm
Ia}$ and supernova rate data used in our analysis are at $z \lesssim 1$.
Therefore the dependence of our results on $\beta$ and $z_{\rm break}$
is not large (see \S \ref{sec:other_parameters}).

\subsection{Delay Time Distribution}
\label{sec:DTD}

To test a variety of DTD of SNe Ia, we use the theoretical models
constructed by \citet{Greggio05} for a wide variety of the progenitor
models (single or double degenerate, and others).  We use four
representative models; two of them are based on the single degenerate
scenario with two different distributions of the secondary stellar
masses adopted in \citet{Greggio05} or \citet{Greggio83}, which are
labeled as ``SD-G05'' and ``SD-GR83'', respectively. The other two
models are based on the double degenerate scenario with two different
treatments of the common envelope phase, which are labeled as
``close-DD'' and ``wide-DD'', respectively.

\begin{figure}
 \begin{center}
  \FigureFile(70mm,40mm){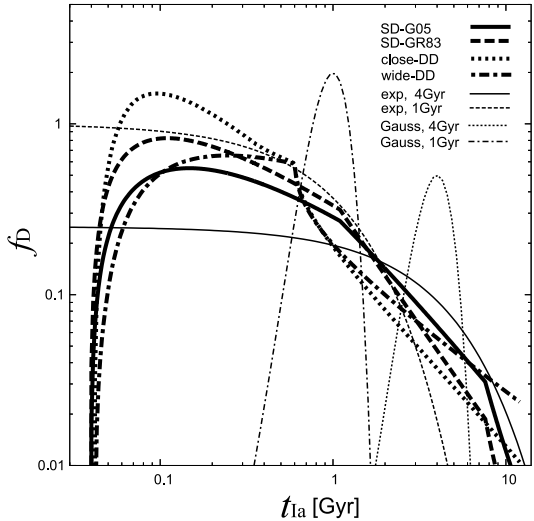} 
 \end{center}
 \caption{ The delay time distribution function for the type Ia supernova
 used in this paper. Thick lines show the
 delay time distribution derived in \citet{Greggio05}, and 
 thin lines show the Gaussian ant exponential models. 
 \label{fig:delay}}
\end{figure}

In addition to these models based on the stellar evolution theory and SN
Ia progenitor scenarios, we also test more phenomenological DTD models
based on simple functional forms as frequently used in SN rate studies
\citep{Madau98, Strolger04}.  One is the Gaussian distribution,
\begin{equation}
f_{D}(t_{\rm Ia}, \tau_{\rm Ia}) = 
\frac{1}{\sqrt{2\pi} (0.2\tau_{\rm Ia}) } 
\exp \left[ - \frac{(t_{\rm Ia}-\tau_{\rm Ia})^2}{2 (0.2\tau_{\rm
Ia})^2} \right],
\end{equation}
and the other is an exponential distribution,
\begin{equation}\label{eq:exp}
f_{D}(t_{\rm Ia}, \tau_{\rm Ia}) = \frac{\exp(-t_{\rm
 Ia}/ \tau_{\rm Ia})}{\tau_{\rm Ia}}. 
\end{equation}
All of the above DTD models are shown in Fig. \ref{fig:delay}.

Recent observations about the dependence of supernova rate on the host
galaxy properties (e.g., galaxy type, stellar mass, star formation, radio
activity) provide evidences for a significant population of SNe Ia whose
rate is directly proportional to the star formation activity (Dallaporta
1973; Oemler \& Tinsley 1979; Della Valle et al. 1994, 2005; Mannucci et
al. 2005; Scannapieco \& Bildsten 2005; Sullivan et al. 2006; Aubourg et
al. 2007). Especially, the correlation with radio galaxies may indicate a
bimordal DTD by two distinct populations (Mannucci et al. 2005). To
test this possibility, we assume a bimodal delay time distribution which
contains the prompt and tardy populations. For the tardy populations we
use the DTD models described above, and the combined DTD with the prompt
population becomes
\begin{equation}
f_{D}(t_{\rm Ia}) = \epsilon_{\rm CSFH} \delta(t_{\rm Ia}) + 
(1 - \epsilon_{\rm CSFH}) f_{\rm delay}(t_{\rm Ia}).
\end{equation}
Here, the tardy part of $f_{\rm delay}(t_{\rm Ia})$ is normalized to the
unity when it is integrated over $t_{\rm Ia}$.  We set $\epsilon_{\rm
CSFH} = 0.4$, which has been inferred from observations (e.g., Sullivan
et al. 2006).  As we will find later, constraints derived by our
analysis are mainly for CSFH, and the DTD modeling does not significantly
affect our main conclusions. In fact, we find that our conclusions are
not significantly changed if $\epsilon_{\rm CSFH} \lesssim 0.7$, and
hence the possible existence of the prompt population is not important
in this work.

\subsection{Comparison with the Data}
The observed data set with which we compare our theoretical model
includes (i) variability magnitudes, redshift, and SN type information
of the GOODS supernova survey \citep{Strolger04} and a supernova survey
in the Subaru Deep Field (SDF-SNS, Poznanski et al. 2007), (ii)
variability magnitude distribution of SSS, and (iii) various SN rate
density data at $z=0$ as well as high-$z$ as tabulated in Table
\ref{table:SNrate}. The majority of the rate density data are for SNe
Ia.

While the shapes of CSFH or DTD have been modeled as above, we treat the
overall normalizations of the SN rate density evolution as free
parameters, separately for Ia and CC SNe. (See \S\ref{sec:MLA} for more
details of the likelihood method.)  Therefore our analysis is free from
the uncertainties in converting star formation rate into supernova rate,
such as the initial mass function, mass ranges of SN progenitors, or
white dwarf explosion efficiency.

\begin{table}
 \begin{center}
  \caption{ SN rate density data \label{table:SNrate}
  }
  \begin{tabular}{ccc}
   \hline 
   z &
   SN rate density&
   Reference \\
   & \footnotesize{$[10^{-4} \rm Mpc^{-3}$ yr$^{-1}$]} & \\ 
   \hline
   \multicolumn{3}{l}{Data for SNe Ia}\\
   0.01 & 0.28 $\pm$ 0.05\footnotemark[$*$] & {Cappellaro et al. (1999)} \\
   0.1  & 0.24 $\pm$ 0.12\footnotemark[$*$] & {Madgwick et al. (2003)}\\
   0.13 & 0.16 $\pm$ 0.07\footnotemark[$*$] & {Blanc et al. (2004)}\\
   0.25 & 0.17 $\pm 0.17$ & {Barris \& Tonry (2006)}\\
   0.30 & 0.34 $\pm$ 0.15 & {Botticella et al. (2007)}\\
   0.35 & 0.53 $\pm 0.24$ & {Barris \& Tonry (2006)}\\
   0.45 & 0.73 $\pm 0.24$ & {Barris \& Tonry (2006)}\\
   0.47 & 0.42 $\pm 0.06$ & {Neill et al. (2006)}\\
   0.5  & 0.48 $\pm 0.17$ & {Tonry et al. (2003)}\\
   0.55 & 0.54 $\pm 0.10$ & {Pain et al. (2002)}\\
   0.55 & 2.04 $\pm 0.38$ & {Barris \& Tonry (2006)}\\
   0.65 & 1.49 $\pm 0.31$ & {Barris \& Tonry (2006)}\\
   0.75 & 1.78 $\pm 0.34$ & {Barris \& Tonry (2006)}\\
   \\
   \multicolumn{3}{l}{Data for CC SNe} \\
   0.01 & 0.43 $\pm$ 0.17\footnotemark[$*$]& {Cappellaro et al.} (1999)
   \\
   \hline
   \multicolumn{3}{@{}l@{}}{\hbox to 0pt{\parbox{80mm}{\footnotesize 
   Note: the rate densities are corrected for the cosmological parameters
   used in this work.
   \par\noindent
   \footnotemark[$*$] The original value has been corrected by using a
   more recent estimate of the local B-band luminosity density, $\rho =
   (1.03 + 1.76 z) \times 10^{8} \LO \rm Mpc^{-3}$ \citep{Botticella07}. 
   }\hss}}
  \end{tabular}

 \end{center}
\end{table}

\subsection{Dust Extinction}
\label{sec:DustExtinction} The effect of dust extinction must be taken
into account.  For comparison with the GOODS, SDF-SNS, and SSS data, we
must calculate light curves of various SN types to calculate the
expected detection number as a function of variability
magnitude. Therefore we first introduce the extinction-corrected light
curves of various SN types, and then they are reddened and absorbed by
the two parameters of the mean color excess, $E(B-V)_{\rm CC}$ and
$E(B-V)_{\rm Ia}$ for CC and Ia SNe, respectively.  We apply a
typical Galactic extinction curve of \citet{Cardelli89} as the
extinction law of both CC SNe and SNe Ia.  Some observations indicate
that the extinction law of low-$z$ SNe Ia might be different from the
Galactic law \citep{Altavilla04, Reindl05, Wang05, Elias-Rosa06}, but it
is still highly uncertain. We confirmed that our results are hardly
changed even when the extinction curve of the Small Magellanic Cloud
\citep{Gordon03} is used. It should be noted that the Calzetti law
(Calzetti et al. 2000), which is often used in studies of high-$z$
galaxies, is an empirical law for the effective attenuation (rather than
extinction) of flux from a whole galaxy, and is not appropriate for
the extinction of flux of a source in a galaxy.

According to the observations of SNe Ia for the cosmological purpose,
the degree of reddening for high redshift SNe Ia seems to be similar to
those of local SNe Ia \citep{Knop03}.  Thus, we assume $E(B-V)_{\rm Ia}
= 0.05$, which is a typical for local SNe Ia \citep{Altavilla04,
Reindl05}, for SNe Ia in all redshifts.  This value of $E(B-V)_{\rm Ia}$
is similar to those used in other SN rate studies. Therefore, we use the
reported values of SN Ia rate densities shown in Table
\ref{table:SNrate} in our likelihood analysis.  One may expect that the
prompt SN Ia population may suffer heavier extinction because generally
star forming regions are dusty. However, the inferred time scale of the
prompt SN Ia events elapsed from star formation is $\sim 10^{8}$ yr, which
is much larger than the lifetime of massive stars leading to CC SNe.

In contrast, CC SNe could suffer from heavier extinction by dust, and it
is also reasonable that the degree of extinction evolves with redshifts
reflecting galaxy evolution. We treat $E(B-V)_{\rm CC}$ as a free
parameter for all high-$z$ CC SNe in the analysis of the GOODS, SDF-SNS,
and SSS data.  On the other hand, we include the local CC SN rate
density of \citet{Cappellaro99} in our likelihood analysis. Comparing
the CC SN light curves used in \citet{Cappellaro99} with those
unreddened, the extinction implicitly included in this estimate is
$E(B-V) \sim 0.2$. Therefore, if we set $E(B-V)_{\rm CC} \sim 0.2$, it
means that there is no evolution for the mean extinction of CC
SNe. Instead, if we get a higher value of $E(B-V)_{\rm CC}$ by the
likelihood analysis, it means that a heavier mean extinction of CC SNe
at high redshifts than in the local universe is required.  We do not
include other CC SN rate density data (e.g., Botticella et al. 2007)
than that of Cappellaro et al., because it is difficult to test
evolutionary models of $E(B-V)$ by an analysis including various rate
density data at different redshifts that are already corrected
for dust extinction with different assumptions.

What we can constrain from the rate density evolution of CC SNe from
$z=0$ to $\sim 1$ is the combination of $\alpha$ and $E(B-V)_{\rm
CC}$. On the other hand, SN Ia rate evolution is affected by $\alpha$.
Therefore, by a combined analysis of CC and Ia SN data, it is expected
to be possible to constrain both $\alpha$ and $E(B-V)_{\rm CC}$, if the
dependence on DTD is weak. We will show that it is indeed possible.

\subsection{The Maximum Likelihood Analysis}
\label{sec:MLA}

Given the theoretical model of SN rate evolution described in the
previous section, we can calculate the expected distribution of the
variability flux and redshift for each SN type, 
$d^2N/(dm_{\rm var} dz)$.  This quantity is calculated by the model
described in OT05 for a given observation filter, time separation, and
detection efficiency of GOODS, SDF-SNS, and SSS data, taking into
account a variety of SN light curve templates and colors.

For the GOODS and SDF-SNS, all the information of $m_{\rm var}$, SN
types, and redshift are available.  Therefore we use the likelihood
function of all the information as:
\begin{equation} 
\label{eq:lhf_gds}
\mathcal{L}_{\rm GOODS/SDF\textrm{-}\rm SNS} = \sum_{j} 
\sum^{N_{\rm obs}^{j}}_{i=1} \ln \left[ \frac {d^2 N_{j}(m_{{\rm
var}, i}, z_i)}{dm_{\rm var}dz} \right] -
N_{\rm exp} \ ,
\end{equation}
where the subscript $i$ is for each supernova, and 
$j$ denotes the SN types (Ia or CC). The total expectation
number $N_{\exp}$ is given by
\begin{eqnarray}
N_{\exp} = \sum_{j} \int dz \int dm_{\rm var} \frac {d^2 N_{j}(m_{\rm
 var}, z)}{dm_{\rm var}dz}\ .
\end{eqnarray}
Note that the likelihood function ${\cal L} \equiv \ln L$ is logarithm
of the likelihood probability $L$ with an arbitrary additive constant. 
(See, e.g., Loredo \& Lamb 1989 for the derivation of
the likelihood function.)

For the SSS data, the type and redshift information is not 
available, and these are integrated out as:
\begin{equation}
\label{eq:lhf_sbr}
\mathcal{L}_{SSS} = 
\sum^{N_{\rm obs}}_{i=1} \ln \left[ \frac {d N_{\rm all}(m_{{\rm var}, i})}
{dm_{\rm var}} \right] - N_{\rm exp} \ .
\end{equation}

For SN rate densities in the literature, 
the likelihood is simply calculated as:
\begin{equation}
\mathcal{L}_{ r(z)} = - \frac{1}{2} \sum_{i} 
\left[ \frac{ r_{\rm i}(z_{i}) - 
r_{\rm model}(z_{\rm i})}{ \sigma_{i}} \right]^{2} \ ,
\end{equation}
where, $r_{\rm model}$ is SN rate density calculated by our model, and
$r_i$, $z_i$ and $\sigma_i$ are observed SN rate
density, mean redshift, and the statistical errors, respectively, of the
$i$-th data point.  These values are summarized in Table
\ref{table:SNrate}, after corrected for the cosmological parameters into
those used in this paper.  As mentioned in \S \ref{sec:DustExtinction},
we use all the rate density data of SNe Ia, but we use only the local
rate density of \citet{Cappellaro99} for CC SNe.

Then, the combined likelihood function for all the data set is given by:
$ \mathcal{L}_{\rm total} =\mathcal{L}_{\rm
GOODS} + \mathcal{L}_{\rm SDF\textrm{-}\rm SNS} 
+ \mathcal{L}_{\rm SSS} +  \mathcal{L}_{ r(z)}
$.  We derive the confidence levels in a standard manner assuming that
$- 2 {\cal L}$ obeys to the $\chi^2$ distribution, and the confidence
regions are determined by the contours of $\Delta \chi^2 = \chi^2 -
\chi^2_{\rm min}$ (e.g., Press et al. 1992).

\section{Constraints on Parameters}\label{sec:results}

\subsection{Constraints on $\alpha$ and $E(B-V)_{\rm CC}$}

\begin{figure*}
 \begin{center}
  \FigureFile(160mm,40mm){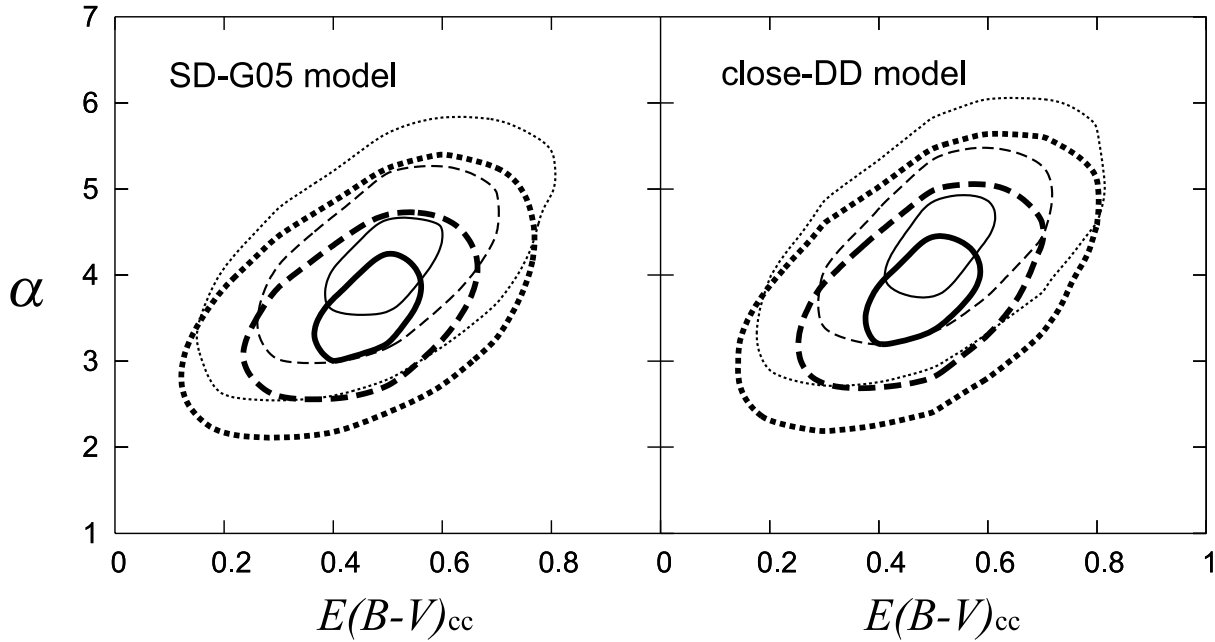}
 \end{center}
 \caption{Confidence regions of the two parameters, $\alpha$ and
 $E(B-V)_{\rm CC}$ derived with the SD-G05 (left panel) and close-DD
 (right panel) models of DTD, respectively.  Confidence contours with
 solid, dotted and dashed curves indicate 68, 95 and 99.7 \% (i.e., 1, 2
 and 3$\sigma$) confidence levels. Thick-lined contours are those
 derived using only off-center SNe in the SSS sample, while the
 thin-lined ones are using both on-center and off-center SNe.
 \label{fig:conf_region}}
\end{figure*}

First we present the confidence regions for $\alpha$ and $E(B-V)_{\rm
CC}$, by using the baseline model, i.e., $\beta = 0$, $z_{\rm break} =
1.5$, and the SD-G05 DTD model (left panel of
Fig. \ref{fig:conf_region}).  In the next subsection we will show that
the constraints on these parameters are robust against the other
parameters of $\beta$, $z_{\rm break}$, and DTD models within the
reasonable ranges. The maximum likelihood is obtained at [$\alpha$,
$E(B-V)_{\rm CC}$] = [$3.7 \pm 0.5$, $0.48 \pm 0.07$] when we use only
the off-center SSS data, while at [$4.2 \pm 0.5$, $0.48 \pm 0.07$] for
the case of using both on- and off-center SSS data. The statistical
errors of these parameters are calculated by one parameter fitting with
the other parameter marginalized. As discussed in \S
\ref{sec:SNsamples}, these two cases should be regarded as the minimum
and maximum numbers of on-center supernovae in SSS.  However, the
difference between these two in this plot is not significant compared
with the statistical uncertainties.  We will show the results derived
from the off-center SSS data as the baseline results in this paper.

Our result indicates $\alpha \sim $ 4.0, and also that there is a
considerable evolution of mean extinction of CC SNe from $z=0$ to $z
\sim 0.5$ (a typical redshift for CC SNe used in the likelihood
analysis).  It should be noted that the constraint on $\alpha$ is
derived only using supernova data sets, without any information on the
other observational estimates of CSFH. Therefore it is interesting to
compare our result with CSFH estimates by galaxy surveys. The estimates
of $\alpha$ based on UV luminosity density are $\alpha = 1.7 \pm 1.0$
\citep{Wilson02} and $\alpha = 2.5 \pm 0.7$ \citep{Schiminovich05}
without taking into account the evolution of extinction. On the other
hand, the estimates based on H$\alpha$ luminosity, which is less
sensitive to the dust extinction, are $\alpha \sim 3.5$ \citep{Tresse02}
and $\alpha = 3.1$ \citep{Doherty06}.  Furthermore, an estimate based
on mid-infrared luminosity gives $\alpha = 4.0 \pm 0.2$
\citep{Perez-Gonzalez05}. Therefore, there is a clear trend that the
estimate of $\alpha$ becomes smaller for methods that are more seriously
affected by dust extinction. (See also Takeuchi et al. 2005.) This trend
is nicely consistent with our result; the true SFR evolution to $z \sim
1$ is described by $\alpha \gtrsim 3$, and small $\alpha$ values
inferred from UV-based estimates can be accounted for by the increase of
mean extinction of star forming regions from $z = 0$ to $z \sim 0.5$.

\begin{table*}
 \begin{center}
  \caption{ Constraints on $\alpha$ and $E(B-V)_{\rm CC}$ and
  thier dependence on $z_{\rm peak}$ and $\beta$ 
  \label{table:zpeak_beta}
  }
  \begin{tabular}{lccc}
  \hline
   & $\alpha$ \footnotemark[$*$]
   & $E(B-V)_{\rm CC}$ \footnotemark[$*$]& 
   Likelihood Ratio \footnotemark[$\dagger$]
   \\
   \hline
   \multicolumn{4}{l}{The baseline CSFH model [($z_{\rm peak}$, $\beta$) = (1.5, 0.0)]}\\
   off-center SSS & 3.7$^{+0.5}_{-0.5}$ & 0.48$^{+0.06}_{-0.07}$ & 1.0\\
   on \& off-center SSS & 4.2$^{+0.5}_{-0.5}$ & 0.48$^{+0.07}_{-0.06}$ & 1.0\\
   \\
   \multicolumn{4}{l}{($z_{\rm peak}$, $\beta$) = (1.2, 0.0)}\\
   off-center SSS & 4.0$^{+0.4}_{-0.4}$ & 0.49$^{+0.07}_{-0.07}$ & 4.1\\
   on \& off-center & 4.6$^{+0.5}_{-0.5}$ & 0.51$^{+0.07}_{-0.07}$ & 5.0\\
   \\
   \multicolumn{4}{l}{($z_{\rm peak}$, $\beta$) = (1.8, 0.0)}\\
   off-center SSS & 3.5$^{+0.4}_{-0.4}$ & 0.47$^{+0.07}_{-0.07}$ & 0.5\\
   on \& off-center SSS & 3.9$^{+0.4}_{-0.4}$ & 0.50$^{+0.07}_{-0.07}$ & 0.7\\
   \\
   \multicolumn{4}{l}{($z_{\rm peak}$, $\beta$) = (1.5, -2.0)}\\
   off-center SSS & 3.7$^{+0.4}_{-0.4}$ & 0.47$^{+0.07}_{-0.07}$ & 1.2\\
   on \& off-center SSS & 4.3$^{+0.5}_{-0.5}$ & 0.49$^{+0.07}_{-0.07}$ & 1.5\\
   \hline
   \multicolumn{4}{@{}l@{}}{\hbox to 0pt{\parbox{105mm}{\footnotesize 
   \footnotemark[$*$] Errors are statistical 1$\sigma$ when
   one of $\alpha$ or $E(B-V)_{\rm CC}$ is marginalized.
   \par\noindent
   \footnotemark[$\dagger$] The ratio of likelihood $L=\exp(\cal L)$ 
   is shown, which is normalized by the value of the baseline CSFH model.  
   }\hss}}
  \end{tabular}
 \end{center}
\end{table*}

\subsection{Dependence on the Other Parameters}
\label{sec:other_parameters} 

In order to show the robustness of our results against variation of the
CSFH model parameters other than $\alpha$, the constraints derived with
different values of $z_{\rm peak}$ and $\beta$, as well as the
likelihood ratio to the baseline model, are summarized in Table
\ref{table:zpeak_beta}. The constraints are almost insensitive to the
high-$z$ CSFH index, $\beta$, within the reasonable range inferred from
galaxy surveys ($\beta \sim -2$--0.) This is because the DTD models
based on the stellar evolution theory have their peak at a relatively
short time scale of $\sim 10^8$ yr. The contribution from the prompt SN Ia
population makes the effect of changing $\beta$ even smaller.  The
constraints do not change significantly within the range of $z_{\rm
peak} \sim$ 1.2--1.8.

To check the dependence on the DTD models, we show the same analysis as
above but using the close-DD DTD model in the right panel of
Fig. \ref{fig:conf_region}. We also show the changes of the constraints
when the DTD model is changed from that in the baseline model in Table
\ref{table:best-fit}.  It can be seen that the derived parameters and
the maximum likelihood for various DTD models are very similar,
indicating that we can derive a robust constraint on $\alpha$ and
$E(B-V)_{\rm CC}$. Though $\alpha \lesssim 3$ is inferred in the cases
of the exponential or Gaussian DTD with a characteristic time scale less
than 1 Gyr, such DTDs are rather unlikely, since these cannot explain
the fact that SNe Ia occur even in the local elliptical galaxies. On the
other hand, we cannot derive any strong constraint on the DTD models,
which is consistent with the result of \citet{Forster06} based on the
GOODS data set, while our analysis includes all available data obtained
in other SN surveys.

\begin{table*}
 \begin{center}
  \caption{ Constraints on $\alpha$ and $E(B-V)_{\rm CC}$ and
  maximum likelihood for different DTD models
  \label{table:best-fit}
  }
  \begin{tabular}{lccc}
  \hline
   & $\alpha$ \footnotemark[$*$]
   & $E(B-V)_{\rm CC}$ \footnotemark[$*$]& 
   Likelihood Ratio \footnotemark[$\dagger$]
   \\
   \hline
   SD-G05 (the baseline model) & \\
   off-center SSS & 3.7$^{+0.5}_{-0.5}$ & 0.48$^{+0.06}_{-0.07}$ & 1.0\\
   on \& off-center SSS
       & 4.2$^{+0.5}_{-0.5}$ & 0.48$^{+0.07}_{-0.06}$ & 1.0\\
   \\
   SD-GR83 & \\
   off-center SSS & 4.0$^{+0.5}_{-0.5}$ & 0.50$^{+0.06}_{-0.07}$ & 2.0\\
   on \& off-center SSS & 4.6$^{+0.4}_{-0.4}$ & 0.53$^{+0.07}_{-0.08}$ & 1.6\\
   \\
   close-DD & \\
   off-center SSS & 3.8$^{+0.4}_{-0.4}$ & 0.50$^{+0.06}_{-0.07}$ & 0.7\\
   on \& off-center SSS & 4.5$^{+0.4}_{-0.4}$  & 0.51$^{+0.06}_{-0.07}$  & 0.8\\
   \\
   wide-DD & \\
   off-center SSS & 4.2$^{+0.5}_{-0.6}$ & 0.53$^{+0.07}_{-0.07}$ & 2.0\\
   on \& off-center SSS & 4.7$^{+0.4}_{-0.4}$  & 0.55$^{+0.07}_{-0.07}$ & 2.7\\
   \\
   exponential 4 Gyr model& \\
   off-center SSS & 4.3$^{+0.5}_{-0.6}$ & 0.52$^{+0.07}_{-0.07}$ & 2.7\\
   on \& off-center SSS & 4.7$^{+0.3}_{-0.3}$ & 0.53$^{+0.06}_{-0.05}$ & 5.0\\
   \\
   exponential 1 Gyr model& \\
   off-center SSS & 3.1$^{+0.4}_{-0.4}$ & 0.43$^{+0.07}_{-0.07}$ & 0.7\\
   on \& off-center SSS & 3.5$^{+0.4}_{-0.5}$ & 0.42$^{+0.07}_{-0.07}$ & 1.2\\
   \\
   Gaussian 4 Gyr model&\\
   off-center SSS & 4.4$^{+0.5}_{-0.6}$ & 0.52$^{+0.07}_{-0.07}$ & 0.1\\
   on \& off-center SSS & 4.8$^{+0.5}_{-0.5}$ & 0.56$^{+0.06}_{-0.06}$ & 0.01\\
   \\
   Gaussian 1 Gyr model& \\
   off-center SSS & 3.0$^{+0.4}_{-0.4}$ & 0.42$^{+0.07}_{-0.07}$ & 0.8\\
   on \& off-center SSS & 3.2$^{+0.4}_{-0.4}$ & 0.42$^{+0.08}_{-0.07}$ & 0.5\\
   \\
   \hline
   \multicolumn{4}{@{}l@{}}{\hbox to 0pt{\parbox{120mm}{\footnotesize 
   \footnotemark[$*$] Errors are statistical 1$\sigma$ when
   one of $\alpha$ or $E(B-V)_{\rm CC}$ is marginalized.
   \par\noindent
   \footnotemark[$\dagger$] The ratio of likelihood $L=\exp(\cal L)$ is
   shown, which is normalized by the value of the baseline model.
   }\hss}}
  \end{tabular}
 \end{center}
\end{table*}

\subsection{On the Type Classification Uncertainties}

It should be noted that the type determination of the GOODS and SDF-SNS
is assumed to be perfectly correct in our analysis. However, SN type
determination is not an easy task; in fact, the ``Bronze'' sample
defined in the GOODS (15 out of all 42 SNe) and 
SNe with intermediate scores of type Ia probability in the SDF-SNS (11
out of all 33 SNe) are thought to be the samples with highly uncertain
type classification. To check the uncertainties about this, we analyze
the GOODS and SDF-SNS data without using the type information of these
SNe. Although the confidence regions are slightly expanded, we obtain
almost the same best-fit parameters of $\alpha$ and $E(B-V)_{\rm CC}$ as
those in our baseline analysis.  Therefore our results are robust
against the type classification in the GOODS and SDF-SNS data set.

\subsection{Cosmic SN Rate Density Evolution}
\label{sec:SNrate}

\begin{figure}
 \begin{center}
  \FigureFile(80mm, 70mm){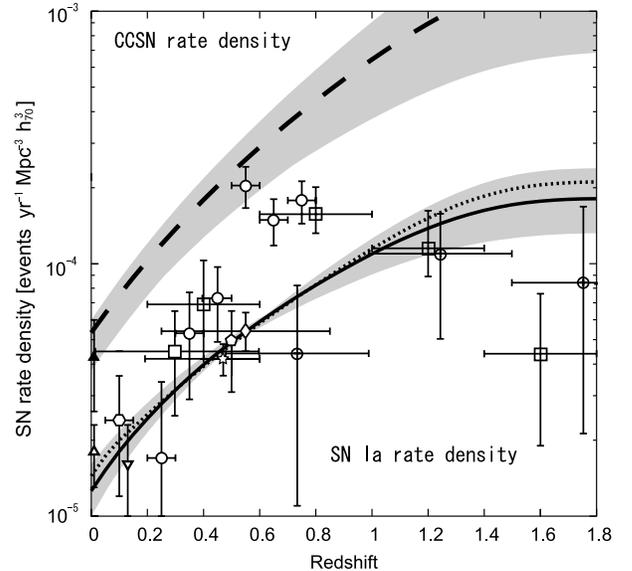} 
 \end{center}
 \caption{
 Time evolution of SN rate density.  The solid and dashed lines are our
 predictions for Ia and CC SNe, calculated from the best-fit parameters
 in our baseline model. Shaded
 regions indicate the deviation within the 1$\sigma$ confidence regions
 of $\alpha$ and $E(B-V)_{\rm CC}$ shown in
 Fig. \ref{fig:conf_region}. The dotted line is the same as the solid
 line, but using the wide-DD DTD model instead of the SD-G05 model.  For
 comparison, data points of SN Ia rate densities used in the likelihood
 analysis are shown in open symbols: Pain et al. (2002, diamond), Tonry
 et al. (2003, pentagon), Madgwick et al. (2003, hexagon), Blanc et
 al. (2004, upside-down triangle), Barris \& Tonry (2006, circle), Neill
 et al. (2006, star), and Botticella et al. (2007, square).  The local
 SN rate densities for Ia and CC SNe (Cappellaro et al. 1999 but
 corrected as in Botticella et al. 2007) are plotted by open and filled
 triangles, respectively.  We also show Ia rate densities derived from
 GOODS data in Dahlen et al. (2004, square with cross) and SDF-SNS data
 in Poznanski (2007, circle with cross), which were not directly used in
 the likelihood analysis since we used the full redshift and magnitude
 distributions of these data.  High redshift data of CC SN rate density
 are not plotted since they significantly depend on dust extinction
 correction.
 \label{fig:SNrate}}
\end{figure}

Figure \ref{fig:SNrate} shows the best-fit model of the cosmic CC and Ia
SN rate density evolution calculated with the SD-G05 model.  
The best-fit model is in reasonable agreement with the observational
data points, but those of the SN Ia rate density at $z \sim 0.6$--0.8
are significantly higher than the model curve. We investigated various
model parameters within our modeling framework, but we cannot reproduce
such a high rate density. This indicates that, if the high rate density
inferred from the data of Barris and Tonry (2006) and Dahlen et
al. (2004) are real, a simple modeling with smooth CSFH and/or DTD
models normally applied in the literature are not sufficient. However,
other SN Ia rates in the same redshift range [SDF-SNS data and the recent
report based on the Supernova Cosmology Project data and the GOODS data
\citep{Kuznetsova07}] are much lower and consistent with the best-fit
model. Strong conclusions cannot be derived for the moment, and more
data in this redshift range are highly desirable.

\section{Summary and Conclusions}\label{sec:summary}

We performed a comprehensive likelihood analysis of almost all available
data of the cosmic SN rate density evolution both for CC and type Ia
supernovae, to get information on the CSFH and the DTD of SNe Ia.  We
utilized the variability magnitude and redshift distribution of CC 
and Ia SNe of the GOODS and SDF-SNS, and other estimates of SN Ia rate
density at various redshift in the literature.  Furthermore, we added
photometrically found supernova candidates in the past imaging data of
Subaru/Suprime-Cam (Subaru Supernova Survey, SSS) to increase the
statistics. The analysis of the SSS data is newly reported here,
including 157 SN candidates down to $i' \sim 26.0$ in the total survey
area of 1.4 deg$^2$.  61 of the 157 SSS candidates are associated with
host galaxies with significant offsets from galaxy centers, and hence
they are almost certainly supernovae. Though the type and redshift
information is not available for SSS, the total SSS SN counts are useful
to constrain the poorly known CC SN rate evolution, by a combination
with the relatively well determined SN Ia rate evolution.

We have tested a variety of DTD models; some of them are based on the
stellar evolution theory, and others are simple analytic functions often
used in the literature. It is found that most of DTD models are
consistent with the current data set, and hence we cannot set strong
constraint on the type Ia SN progenitor.

On the other hand, this rather week dependence on DTD models is an
advantage when one tries to constrain CSFH parameters.  It is required
that $\alpha$ (the SFR evolution index from $z=0$ to $\sim 1$) is 3--4,
with a considerable evolution of mean extinction of CC SNe, as $E(B-V)
\sim 0.2$ at local and $E(B-V) \sim 0.5$ at $z \sim 0.5$. Since we did
not utilize any information from CSFH estimates by galaxy surveys, we
can compare our result with those by galaxy surveys. Recent estimates
based on UV luminosity are $\alpha \lesssim 2.5$, while those based on
H$\alpha$ or mid-infrared luminosity are close to our result, $\alpha
\sim$ 3--4.  These are nicely consistent with our finding of the
significant evolution of extinction for CC SNe, indicating a strong
evolution of extinction of star formation activity in the universe even
at $z \sim$ 0--0.5.

The consistency between CSFH based on SFR in galaxies and SN rates is
not trivial. Most indicators of galactic SFR trace the production of UV
or ionizing photons and hence the formation of massive stars, which is
the same as CC SNe. However, an evolution of IMF of massive stars
could change the ratio of UV photon production to CC SN rate. Our result
implies a roughly constant production efficiency of SNe per unit mass of
star formation, and this would give some constraint on IMF evolution or
metallicity effects. We have also demonstrated that, based on the
counts of SN candidates without host galaxies, the contribution to the
cosmic star formation activity from faint galaxies under detection limit
or intergalactic star formation should not be significant. This is a
clear advantage of CSFH constraint from supernova surveys, which cannot
be obtained by CSFH studies based on galactic SFR estimates.

The authors would like to thank L. Greggio for providing the data of
delay time distribution and useful discussions. We also appreciate
useful comments from A. Gal-Yam, D. Maoz, D. Poznanski, and an anonymous
referee.  A part of the SSS data was obtained as a part of the Supernova
Cosmology Project (SCP), and we also thank the SCP members for useful
discussions.  This work was supported by the Grant-in-Aid for the 21st
Century COE ``Center for Diversity and Universality in Physics'' from
the Ministry of Education, Culture, Sports, Science and Technology
(MEXT) of Japan. T.O. has been supported by JSPS Research Fellowships
for Young Scientists. T.T., N.Y., and M.D. are supported by the JSPS -
USA bilateral programme.

\end{document}